\documentclass[5p, authoryear,10pt]{elsarticle}

\usepackage{float}
\usepackage{amssymb,amsfonts,amsmath}
\usepackage{setspace}
\newcommand{\beq}{\begin{equation*}}
\newcommand{\eeq}{\end{equation*}}
\newcommand{\beqn}{\begin{equation}}
\newcommand{\eeqn}{\end{equation}}
\newcommand{\vect}[1]{\textbf{#1}}

\usepackage[bookmarksnumbered,colorlinks,backref, bookmarks, breaklinks, linktocpage, linkcolor=blue, citecolor=red]{hyperref}

 \usepackage{lineno}

\begin{document}

\begin{frontmatter}



\title{Implication of the lopsided growth for the viscosity of Earth's inner core}


\author{Hugau Mizzon and Marc Monnereau}

\address{IRAP, University of Toulouse, CNRS, Toulouse, France}

\begin{abstract}
Two main seismic features characterize the Earth's inner core: a North-South polar anisotropy and an East-West asymmetry of P-wave velocity and attenuation. Anisotropy is expected if shear deformation is induced by convective motions. Translation has recently been put forward as an important mode of convection of the inner core. Combined with a simple diffusive grain growth model, this mechanism is able to explain the observed seismic asymmetry, but not the bulk anisotropy. The source of anisotropy has therefore to be sought in the shear motions caused by higher modes of convection. 
Using a hybrid finite-difference spherical harmonics Navier-Stokes solver, we investigate the interplay between translation and convection in a 3D spherical model with permeable boundary conditions at the inner core boundary. Three parameters act independently: viscosity, internal heating and convection velocity in the outer core. Our numerical simulations show the dominance of pure translation for viscosities of the inner core higher than $10^{20}$ Pas. Translation is almost completely hampered by convective motions for viscosities lower than $10^{18}$ Pas. Between these values, translation and convection develop, but convective downwellings are restricted to the coldest hemisphere where crystallization occurs. On the opposite side, shear is almost absent, thereby allowing grain growth. We propose that the coexistence of translation and convection observed in our numerical models leads to a seismic asymmetry but localizes deformation only in one hemisphere.
\end{abstract}


\end{frontmatter}


\newpage
\section*{Introduction}

The image of the inner core growing slowly at the center of the Earth by gradual cooling and solidification of the surrounding liquid outer core is being replaced by the more vigorous image of a ``deep foundry'' \citep{Buffett2011}, where melting and crystallization rates exceed by many times the net growth rate \citep{Monnereau2010, Alboussiere2010,Gubbins2011}

During seventy years,  the analysis of compressive waves (P waves) and free oscillations excited after large earthquakes have been depicting a more and more complex structure of the inner core. It appears anisotropic, with a fast axis parallel to Earth's spin axis. It is also asymmetric: within the outermost 100 km, velocity and attenuation of P waves increase from the hemisphere facing America (West) to the one facing Asia (East) \citep{Tanaka1997} ; anisotropy also seems stronger in the Western hemisphere than in the Eastern one \citep{Deuss2010}.

In the 80's and 90's, anisotropy was thought to be the prominent feature. This is commonly attributed to the preferred orientation of iron crystals, possibly acquired during solidification but most probably resulting from creep flow. Thermal convection, developing a flow characterized by a spherical harmonic degree $l=1$ (a clementine shape), orientated along the spin axis of the Earth, was one of the first candidates to account for anisotropy \citep{Jeanloz1988}. In this model and the following ones \citep{Weber1992, Buffett2009}, the nature of the inner core boundary (i.e. a phase change) was not considered, a classical impermeable boundary condition being preferred. With permeable boundary conditions --- a phase change does not prevent material transfer ---, the expression of convection at the first harmonic degree is a constant velocity field across the inner core, that obviously does not produce any deformation, but implies melting on one side and crystallization on the opposite side. This peculiar situation, called translation of the inner core, has recently been put forward to explain the hemispherical asymmetry of velocity and attenuation \citep{Monnereau2010}.

 If the inner core grows in a superadiabatic regime, which is the condition for the onset of convection, an unstable thermal stratification develops, so that any infinitesimal thermal heterogeneity of harmonic degree $l=1$ (i.e. one side colder than the other one) will be amplified. Such a heterogeneity will induce a displacement of the inner core to maintain its center of mass --- shifted toward the denser/colder hemisphere --- at the center of the Earth. The inner core acquires a positive topography on the hotter and lighter side and a negative one on the opposite side. This topography is thermodynamically unstable: the side emerging above the phase change melts, bringing hotter material up to the surface, while the sinking side allows crystallization. The phase change acts to remove the topography, which is continuously rebuilt by isostatic equilibrium. This feedback results in a permanent drift from the crystallizing side to the melting side, the drift velocity being controlled by the ability of the outer core to restore the adiabatic condition at the surface of the inner core. 

Translation is consistent with multiple scattering models of wave propagation. If they do not experience deformation, iron crystals grow as they transit from one hemisphere to the other. Larger crystals constituting a faster and more attenuating medium, a translation velocity of some cm/yr (about ten times the growth rate) is enough to account for the superficial asymmetry observed for P-wave velocity and attenuation, with grains of a few hundred meters on the crystallizing side (West) growing up to a few kilometers before melting on the East side, and a drift direction located in the equatorial plane.

Translation was also proposed to be responsible for the formation of a dense layer at the bottom of the outer core, since the high rate of melting and crystallization would release a liquid depleted in light elements at the surface of the inner core \citep{Alboussiere2010}. This would explain the anomalously low gradient of P wave velocity in the lowermost 200 km of the outer core \citep{Poupinet1983}.

Clearly, translation cannot account for anisotropy. However, convective modes developing at higher harmonic degrees ($l > 1$) will necessarily induce shear deformation that could be a source of anisotropy. The development of these modes depends on the Rayleigh number, that controls the vigor of the convection, and thus mainly on the viscosity of the solid portion of the core. At high Rayleigh number (low viscosity), these modes can be dominant and dissipate the  degree $l=1$ of the thermal heterogeneities: the source of the translation. Thus a viscosity threshold may be expected below which translation would not take place. This may constrain the viscosity of iron at the conditions of the inner core, based on seismological observation. In this paper, we present dynamics model of inner core taking into account the phase change boundary and study the interaction of translation and convection.

\section*{Model setup}
Inner core dynamics obey similar governing equations to those used for mantle convection. The specificity lies in the boundary conditions required to treat the phase change at its surface.
\subsection*{Momentum equations}
 Inertial forces can be neglected because of the high viscosity of solid iron at the inner core temperature and pressure conditions, which is at least $10^{16}$ Pas \citep{Yoshida1996}.  Conservation of momentum just expresses the balance between buoyancy forces and viscous dissipation. It is time independent. 
\begin{equation}
\nabla \cdot \tau - \nabla p = - \rho \textbf{g}.
\label{Navier}
\end{equation}
 $\tau$ is the deviatoric stress tensor, $p$ the pressure, $\rho$ the density and $\textbf{g}$ the gravity.
The permeable surface condition is introduced. It describes the balance between the radial stress and buoyancy forces induced by topography: 
\begin{equation}
\tau_{rr}-p|_{R_{ic}}=(\rho_l-\rho_s) g_{icb} h,
\label{perm}
\end{equation}
where $\tau_{rr}$ is the deviatoric radial stress, $R_{ic}$ the inner core radius, $\rho_s$ the density of solid iron, $\rho_l$ the density of liquid iron, $g_{icb}$ the gravity at the surface of the inner core and $h$ the topography. In addition, the surface is considered as tangential stress free:
\begin{equation}
\tau_{r\theta, \phi}|_{R_{ic}}=0
\label{fs}
\end{equation}
At the first spherical harmonic degree $l=1$, the momentum equation written for the full sphere with the above conditions comes down to a simple isostasy equilibrium where thermal heterogeneities are balanced by a topography ($\nabla \cdot \tau = 0$ in Eq. \textbf{\ref{Navier}}) \citep{Monnereau2010}. This singularity requires a particular treatment.  
The three components of the first harmonic degree of the topography $h_{1m}$ are directly related to the position of the center of mass anomaly caused by the presence of the inner core in the outer core:

\begin{equation}
h_{1m} = \dfrac{\rho_s \alpha }{\rho_s-\rho_l} \int_0^{R_{ic}}\left(\dfrac{r}{R_{ic}}\right)^3 \Theta_{1m}(r) dr,
\label{isost}
\end{equation}
with $\alpha$ the thermal expansion coefficient and $\Theta_{1m}(r)$ the radial profile of the temperature heterogeneity at degree $l=1$ and order $m$ (see Appendix A for more details).

\subsection*{Surface heat exchange with the outer core}

The topography is thermodynamically unstable and eroded at a rate depending on the vigor of convection within the outer core.  In the mantle, the position of mineral phase transitions is mainly related to the ambient temperature, latent heat exchanges having almost no effect. For instance, a transition with a positive Clapeyron slope like olivine to spinel occurs deeper in ascending (hot) currents than in dipping slabs. The reverse situation happens for the ICB. The turbulent flow in the liquid maintains the temperature above the surface close to the adiabat so that the topography only depends on the temperature variations induced by the latent heat effects and not on temperature anomalies within the inner core. Topography is thus positive where material exits the inner core because of the cooling induced by the latent heat consumption.  The local thermodynamical  equilibrium is achieved  when the rate at which the inner core consumes or releases latent heat equals the rate at which the outer core brings or takes the energy to maintain the adiabatic temperature. This can be written as  \citep{Alboussiere2010}:

\begin{equation}
v_r(R_{ic})= F h,
\label{thermo}
\end{equation}
where 
\begin{equation}
F=\dfrac{-\rho_l g_{icb} \left( \dfrac{\partial T_m}{\partial p} - \dfrac{\alpha T_S}{\rho_l C_p} \right) u_l C_p}{L}.
\end{equation}
$v_r(R_{ic})$ is the radial velocity across the phase change, $T_m$ the melting temperature of iron, $T_S$
 the adiabatic temperature, $u_l$ the amplitude of the outer core convective flow at the surface of the inner core, $C_p$ the specific heat and L the latent heat of the phase change. We studied the kinetics of this equilibrium in a simple 1D model (see Appendix B) in order to check that it can be considered as instantaneous. The temperature adjustment is reached in a few thousands years which is at least one order of magnitude smaller than the characteristic time of the convective fluctuations at the highest Rayleigh number we considered.

Since Eq.\;\textbf{\ref{thermo}} is linear and time independent, it applies in the spectral domain, so that the convection and isostatic problems can be separated. The velocity of translation is directly obtained from $h_{1m}$ (Eq.\;\textbf{\ref{isost}}) and topography at higher degree ($l>1$) is computed through the resolution of Eqs.\;\textbf{\ref{Navier}}-\textbf{\ref{fs}}.

\subsection*{Energy equation}
The inner core dynamics is a moving boundary problem that can be treated in all its complexity \citep{Deguen2011a}. For the sake of simplicity, we may neglect the variation of the radius with time, and focus on the present time dynamics. This assumption is plainly justified since the translation velocity required to account for seismic properties of the inner core, but also for the formation of  a dense layer above ICB, should exceed the inner core growth rate (by one or two orders of magnitude) \citep{Monnereau2010, Alboussiere2010}. The energy equation is thus written in terms of the temperature relative to the adiabat anchored at the ICB ($\Theta=T-T_S$), in which the decrease of ICB temperature with time plays the role of an internal heating:
\begin{equation}
\rho_s C_p \dfrac{D\Theta}{D t} + \alpha \rho_s g \Theta v_r + \tau : \nabla \textbf{v}  + k \Delta \Theta = \Phi,
\label{heat}
\end{equation}
where  $\textbf{v}$ is the velocity vector, $k$ the conductivity, $\Phi$ the internal heating rate and $\Delta$ the Laplacian operator. The surface is assumed isothermal, $\theta=0$. This comes to neglect the small temperature perturbation due to the latent heat ($\sim10^{-3}$K) compared to the one involved in the dynamics ($\sim$1K). Indeed, the latent heat effects play a much more important role in the momentum equation than in the energy equation. 

\subsection*{Parameters}
$\Phi$, the internal heating, results from a competition between the heat lost by conduction along the adiabat and the secular cooling of the core:
\begin{equation}
\Phi=k\Delta T_S - \rho_s C_p \dfrac{d T_S}{d t}.
\label{phi}
\end{equation}

\begin{table*}[t]
\caption{Value of parameters used in the numerical experiments.}
\begin{tabular*}{\hsize}{@{\extracolsep{\fill}}llcl}
 \hline
 Symbol & Name & value  & unit \\
 \hline
$\alpha$ & Thermal expansion coefficient                       & $2.5\times 10^{-5} $  & K$^{-1}$   \\
$C_p$ & Heat capacity                                   & 800                            & J/K kg       \\
$T_{S}$ & Adiabatic temperature anchored at ICB  & 5500                          & K              \\
$k$ & thermal conductivity               & 34                              & W/K m       \\
${\partial T_m}/{\partial p}$ & melting curve slope&            & K/Pa         \\
$L$         & Latent heat                  & 800                            & J/kg           \\
$\rho_s$ & solid iron density                          & 11800                        & kg/m$^3$  \\
$\rho_s$ & liquid iron density                         & 11200                        & kg/m$^3$  \\
$R_{ic}$  & inner core radius                         & 1220                          & km             \\
$g_{icb}$ & gravity at the ICB                        & 4                                & m/s$^{2} $  \\
 \\
$u_l$  & Outer core  velocity  &                   $10^{-5}  -  10^{-3}$                  & m/s     \\
$\Phi$ & Internal heating                         & $2.5\times 10^{-10}  -  \;2.5 \times10^{-8}$  & W/m$^3$  \\
$\eta$ & Viscosity                                     &                 $10^{16}  - 10^{20}$                  & Pa s       \\
$Ra$  & Rayleigh number                        &        $6\times10^{4}  -  6\times10^{9}$        &             \\
\hline
\label{table_const}
\end{tabular*}
\end{table*}

The superadiabatic regime, required for the onset of translation or convection, is defined by $\Phi>0$. Uncertainties on both the conductivity and the cooling rate contribute to the uncertainty on $\Phi$. The former was evaluated around 60\;W/mK \citep{Stacey2001} ; an estimate that has been recently revisited and reduced by a factor of 2: 36\;W/mK  \citep{Stacey2007}. $d T_S/d t$ is proportional to the inner core growth rate $\dot{R}_{ic}$ : 
\begin{equation}
\dfrac{d T_S}{d t}=-\rho_l g_{icb}   \left(\dfrac{\partial T_m}{\partial p} - \dfrac{\alpha T_S}{\rho_l C_p} \right) \dot{R}_{ic},
\end{equation}
$\dot{R}_{ic}$  being itself proportional to the heat flux at the core-mantle boundary (CMB) \citep{Labrosse2003}, whith walues for the latter estimated to be 6\;TW and 14\;TW \citep{Turcotte2002}.  It results that $\Phi$ can be as large as $2.5 \times 10^{-8}$\;W/m$^3$.

The vigor of convection is controlled by the Rayleigh Number: 
\begin{equation}
Ra=\dfrac{\rho_s^2 g_{icb} C_p \alpha R_{ic}^5 \Phi}{6 k^2 \eta},
\end{equation}
written here for pure internal heating in a spherical system. At first sight, the variation of radius appears of primary importance for the evolution of the dynamics of the inner core along its history, notably because of the successive development of convective modes at higher and higher degree as the inner core grows  \citep{Deguen2011a}. However, as mentioned before, we focus on the present day dynamics. In this case, the least constrained parameters are the internal heating, $\Phi$ and the viscosity, $\eta$, both subject to uncertainties of several orders of magnitude.

A minimum viscosity of 10$^{16}$ Pas was deduced from attenuation of seismic waves in the inner core and a maximum of 10$^{21}$ Pas was inferred from conjectures on iron rheology close to its melting point for pressures at the center of the Earth \citep{Yoshida1996}. Since translation does not involve deformation, it is not sensitive to $\eta$, but only to $\Phi$. Their effects have to be studied independently. 

The third and last parameter is the amplitude of the convective velocity within the outer core, $u_l$, that acts only on the  translation velocity, the effect of the topography remaining very small on the convective circulation.  Other parameters in the expression of the factor F (Eq.\;\textbf{\ref{thermo}}) remain much more constrained than $u_l$. The average convection velocity in the outer core was assessed to be 10$^{-4}$ m/s with 10$^{-3}$ m/s maxima, from the secular variations of the geomagnetic field \citep{Amit2006}. It is possible that a dense layer forms at the base of the outer core \citep{Alboussiere2010,  Buffett2011}, convection velocity could be lower at the surface of the inner core. We will also test a velocity ten times lower than the average.

\subsection*{Numerical methods}
The momentum equation is solved using a classical spherical harmonics expansion together with a radial finite difference solver for each harmonic. The conservation of energy is solved with a second order finite volume method. More details on the code may be found in \cite{Monnereau2002b}. For all calculations we have employed 128 spherical harmonics and 128 points in the radial direction.

\section*{Results}

Table\;1 summarizes the range of investigation for the parameters $\Phi$, $\eta$ and $u_l$. Figure\;\ref{fields} presents the velocity fields of experiments with an intermediate value for the outer core velocity, $u_l=10^{-4}$\;m/s. A rapid overview shows that translation dominates at high viscosity and high internal heating (Fig.\;\ref{fields}c, e-i), and is absent from experiments at low viscosity and low internal heating (Fig.\;\ref{fields}a, b \;\&\;d). Translation is characterized by a surface radial velocity, i.e. the radial velocity through the ICB, positive on one hemisphere and negative on the opposite, inducing melting and crystallization respectively.

\subsection*{Viscosity effect}
At high viscosity, $\eta=10^{20}$\;Pas, translation seems to be the sole active mode of heat transfer. Nevertheless, the velocity field does not coincide with a pure solid motion, except at very low internal heating rate (Fig.\;\ref{fields}g). It also contains a weak degree $l=2$ component, attesting the development of convective modes at degree $l>1$. This feature is not visible on the surface radial velocity maps, but is revealed by the non-collinearity of the velocity vectors depicted on the cross section of Fig.\;\ref{fields}h. 
At lower viscosity, $\eta=10^{18}$, translation does not develop for the lowest internal heating rate Fig.\;\ref{fields}d. A degree $l=2$ dominates in this case, where a sheet-like downwelling draws a great circle. When translation still takes place  (Fig.\;\ref{fields}e\;\&\;f), it is associated with one or several planar dipping currents. These short wavelengths structures only form within the crystallizing hemisphere. 
At low viscosity, $\eta=10^{16}$\;Pas, these 2D currents give way to cold spots. Translation disappears at low and intermediate internal heating rates (Fig.\;\ref{fields}a\;\&\;b). It persists at the highest heating rate, but, contrary to the situations at $\eta=10^{18}$\;Pas, high degree convective structures are present on both crystallizing and melting hemispheres even if an asymmetry in the number of structures is still present.

 In summary, a low viscosity enhances the vigor of the convection and favors the development of short wavelength structures, a result expected since it corresponds to an increase in Rayleigh number.

\subsection*{Internal heating effect}

\begin{figure*}[t]
\centerline{\includegraphics[width=18cm]{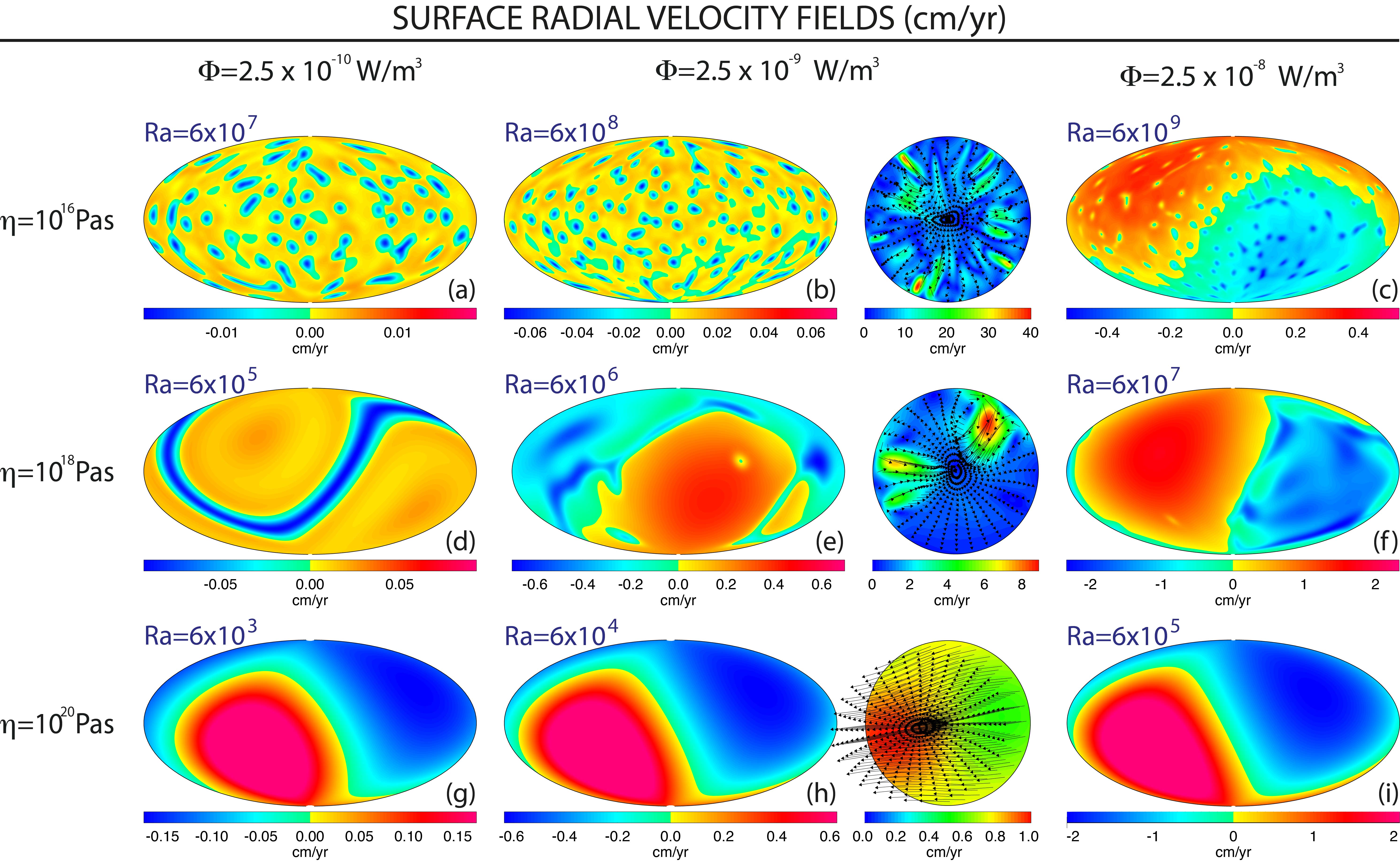}}
\caption{Effects of viscosity and internal heating on the style of convection in the inner core. Surface radial velocity fields (hammer projection) are displayed for experiments with 3 values of viscosity $\eta$ and internal heating $\Phi$. They are snapshots taken after that convection equations were integrated to a quasi steady-state regimes. Equatorial cross-sections of the velocity field are also shown for the intermediate internal heating value $\Phi=2.5\times10^{-9}$\;Wm$^{-3}$.  The background color is function of the velocity amplitude. Note the predominant influence of viscosity, with complex convection modes at low viscosity, simple translation mode at high viscosity and hybrid circulation pattern for intermediate viscosity. The Rayleigh number, Ra, is indicated in blue to the upper left corner of each plot. Identical Rayleigh numbers, corresponding here to identical $\Phi/\eta$ ratio, may result in different convective pattern, because internal heating has a larger impact on translation than on convection.}\label{fields}
\end{figure*}

Internal heating also enhances convection, as evidenced by the increase of the number of structures. However, when degree $l=1$ dominates the circulation, the major effect is an amplification of the translation, whose velocity seems only dependent on the internal heating rate. For instance, the comparison of situations depicted in Fig.\;\ref{fields}e\;\&\;h and Fig.\;\ref{fields}f\;\&\;i. shows that translation reaches the same velocity at $\eta=10^{20}$\;Pas and $\eta=10^{18}$\;Pas, irrespective of the convective pattern. This highlights the fact that the Rayleigh number is not an independent parameter of the coupled system. Besides, this system expresses differently at equal Rayleigh numbers as in Fig.\;2a where point like cold currents inhibit the translation and in Fig.\;2f where the reverse prevails. Internal heating has a promoting effect on the translation.

\subsection*{Translation vs convection}
Translation induces an asymmetry of the thermal boundary layer with a thickening on the crystallizing side and a thinning on the melting side. When this thinning is enough to reduce the local Rayleigh number of the thermal boundary layer below the critical value, no convective instability can begin. This leads to situations where convective structures are restricted to the crystallizing side as in Fig.\;\ref{fields}e\;\&\;f. This also explains the large difference between Fig.\;\ref{fields}d\;\&\;i, both corresponding to experiments with a Rayleigh number more than 200 times the critical value for the onset of convective structure at a spherical harmonic degree $l=2$ (2607 with impermeable surface conditions \citep{Chandrasekhar1961}). At high Rayleigh number, when the conditions of this thinning are not fulfilled, higher harmonic degree structures, such as coldspots, develop. These uniformly distributed downwellings suppress the thermal anomaly at harmonic degree $l=1$ that translation requires to develop. Thus a real competition is playing between convection and translation. 

\begin{figure}[h]
\centerline{\includegraphics[width=7 cm]{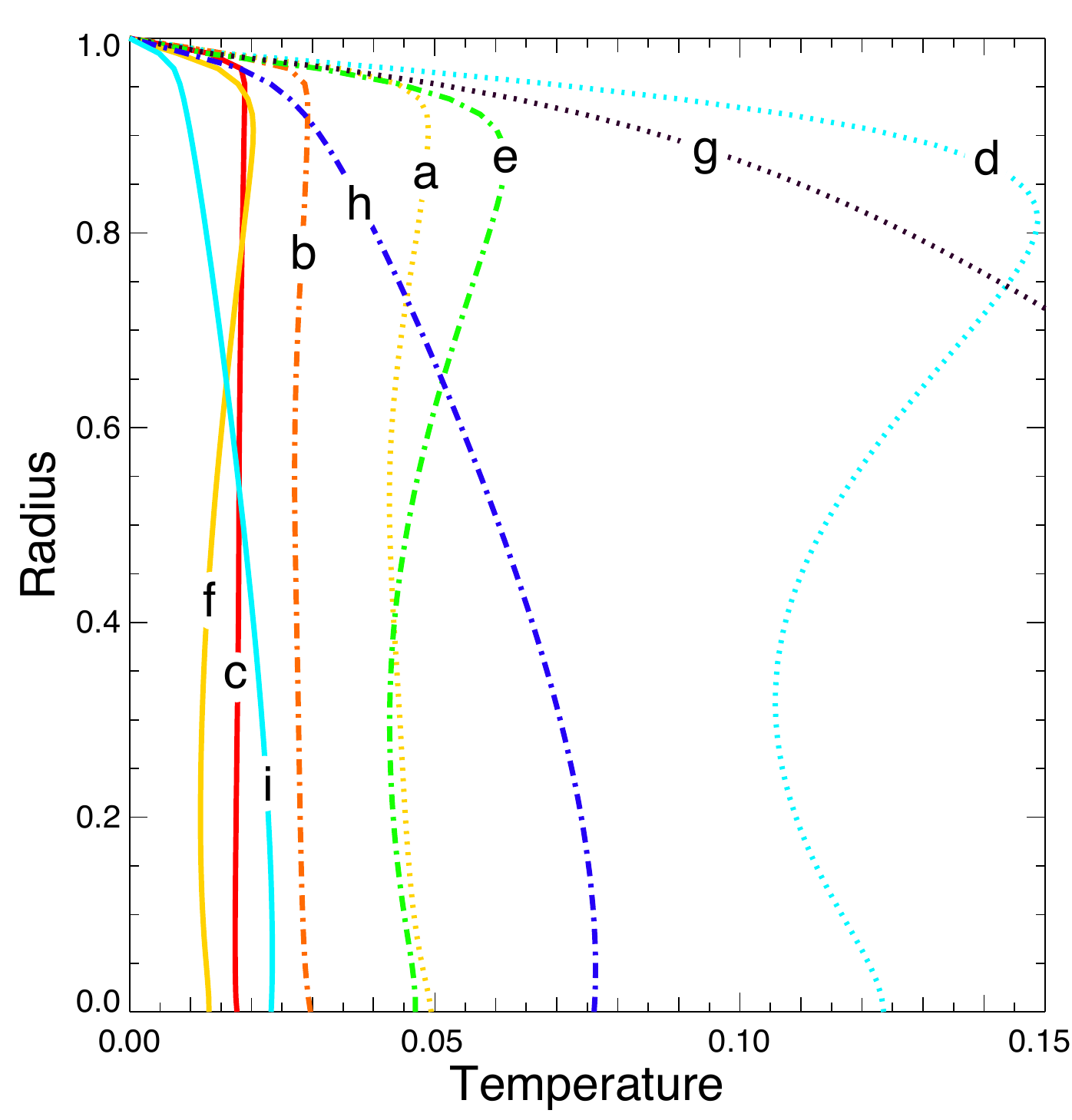}}
\caption{Non dimensional temperature profile of experiments shown in Fig.\;\ref{fields}. Labels are the same as in  Fig.\;\ref{fields}. The non dimensional temperature can be scaled by $\Phi R_{ic}^2/6k$, where the conductivity, $k$, has been set to 36\;W/mK. The line style is related  to the internal heating rate: dotted lines for $\Phi=2.5\times10^{-10}$W/m$^3$, dot-dashed lines for $\Phi=2.5\times10^{-9}$W/m$^3$, plain lines for $\Phi=2.5\times10^{-8}$W/m$^3$. The color changes from dark blue to red as the Rayleigh number increases. Rayleigh numbers are indicated on  Fig.\;\ref{fields}. For instance, profiles d and i, or a and f correspond to experiments with the same Rayleigh number, $6\times10^5$ and  $6\times10^7$  respectively. A comparison of these profiles shows the strong cooling induced by translation. Profiles resulting from almost pure translation are parabolic (i, h \& g) and evolve toward the classical shape observed for convection in a full sphere (a and b).}
\label{temp_prof}
\end{figure}

These are two distinct modes of heat transfer, which can be verified by examining global quantities. For instance, horizontally averaged temperature profiles exhibit very different shapes depending on whether translation operates or not. The classical temperature profile obtained for convection in a full sphere is marked by a top thermal boundary layer and a maximum located just below. When the translation develops, this shape evolves towards an elliptic integral profile $E(\pi,r)$ where the maximum is now reached at the center (see in Fig.\;\ref{temp_prof} changes from profiles b to e, then f). This is the analytical solution of the temperature equation when conduction and time dependence are neglected. An additional major effect is the strong cooling produced by the translation as illustrated by the comparison of profiles of experiments run at the same Rayleigh number (a and f, or d and i). The permeable boundary that allows advective heat transfer across the surface is responsible for this cooling. This result was already shown in a former study, where such permeable conditions were used to mimic the cooling due to intense volcanism on Io \citep{Monnereau2002a}. In this case, the exponent of the power scaling relation between averaged temperature and Rayleigh number is $-1/2$ with permeable conditions, instead of the classical value $-1/4$ found with impermeable conditions \citep{McKenzie1974}. 

\begin{figure}[h]
\centerline{\includegraphics[width=7 cm]{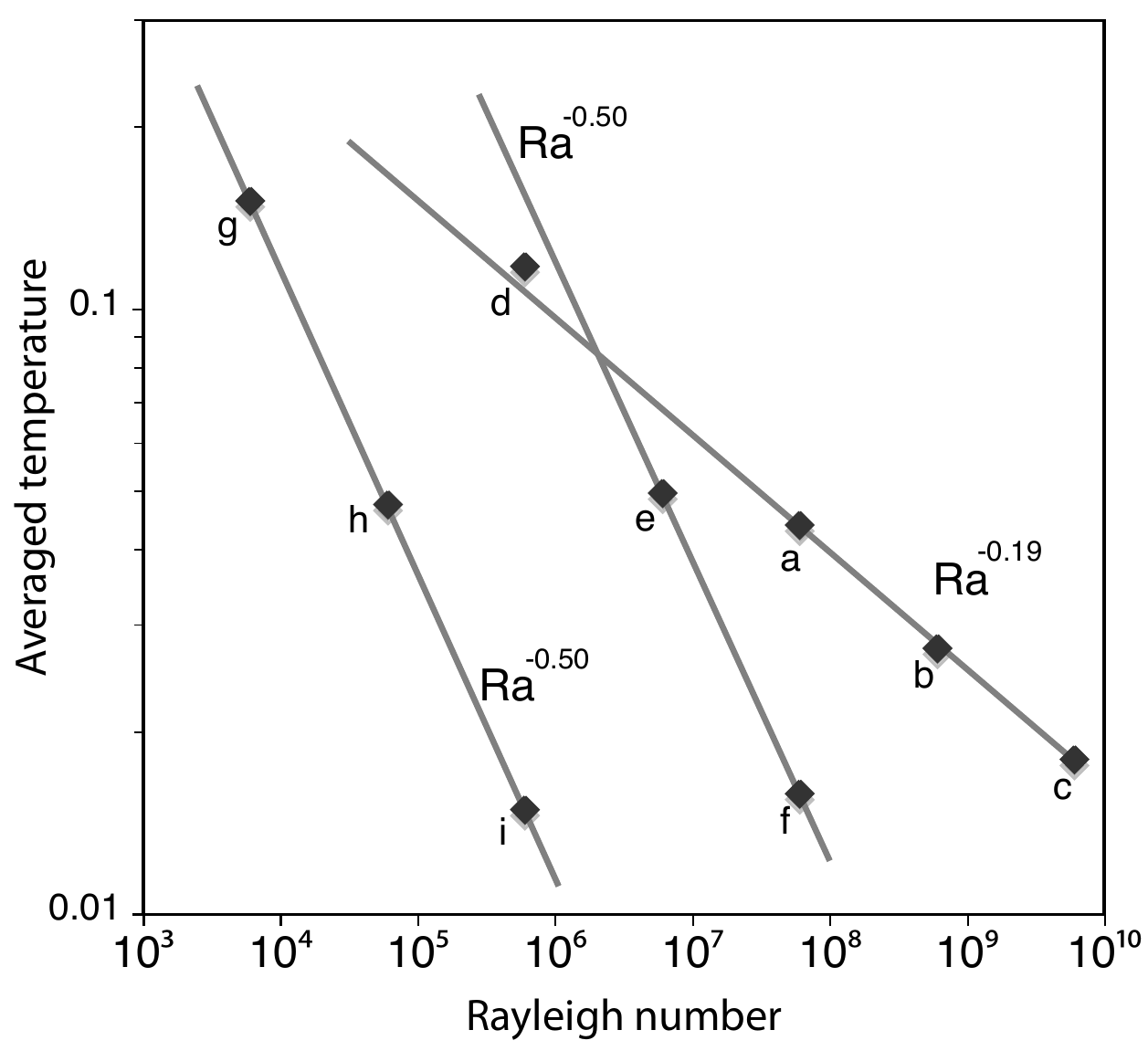}}
\caption{Non dimensional averaged temperature as a function of the Rayleigh number of experiments shown in Fig.\;\ref{fields}. Labels are the same as in  Fig.\;\ref{fields}. }
\label{Tm_Ra}
\end{figure}

The averaged temperature decreases proportionally with $Ra^{-1/2}$ only when translation dominates the circulations (see Fig.\;\ref{Tm_Ra}). Elsewhere the power low exponent is $-0.2$. Actually, the non dimensional averaged temperature is just proportional to $\Phi^{-1/2}$, since viscosity has no effect on translation. This relation can be easily understood by considering that, when translation dominates, the energy equation comes down to a balance between internal heating and advection, so that the product of velocity and temperature is proportional to $\Phi$. This also expresses the fact that all the energy produced internally is evacuated by advection. As the translation velocity is proportional to the temperature (see Eqs \textbf{[\ref{isost}]} \& \textbf{[\ref{thermo}]}), the temperature is proportional to $\Phi^{1/2}$. Finally, the non dimensional temperature is obtained by dividing the temperature by the temperature scale $\Phi R_{ic}^2/6k$. When the geometry does not allow the translation mode, as in the case of convection between spherical shells, or for all convective modes at degree $l>1$, the velocity is proportional to $\Phi/\eta$, so that the non-dimensional temperature varies in proportion to $Ra^{-1/2}$. 

\begin{figure}
\centerline{\includegraphics[width=8.7cm]{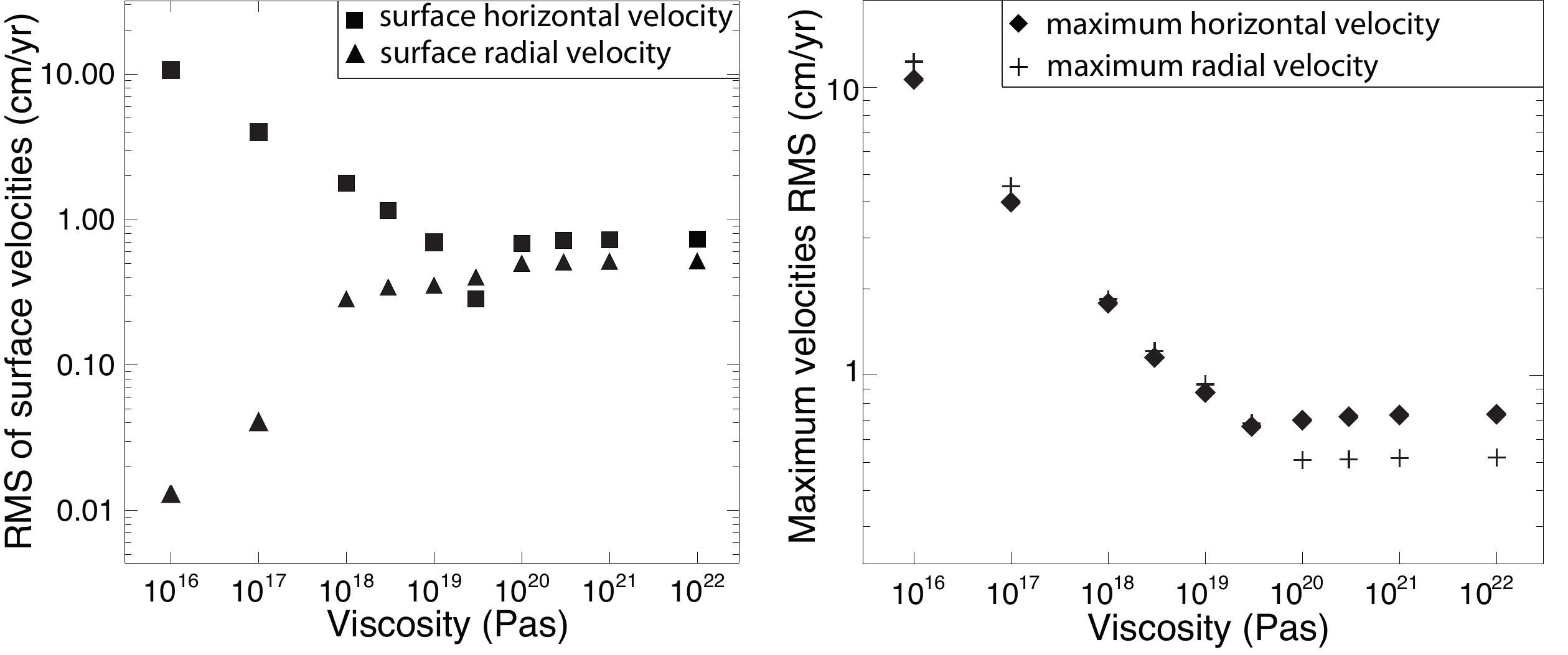}}
\caption{Effect of viscosity on surface velocities (left) and maximum velocities (right). A threshold around $10^{19}$\;Pas is clear, with constant velocities above denoting that solid translation prevails. Below, the surface radial velocity decreases whereas the surface horizontal velocity increases, denoting a change 
 of the ICB from permeable to non-permeable state.The maximum velocity reveals also this change showing a constant exchange with the outer core above $10^{19}$Pas, and a power law dependence characteristic impermeable conditions below (the slope is around -2/5). These experiments have been run with $u_l=10^{-4}$m/s and $\Phi=2.5 \times 10^{-9}$W/m$^3$}
\label{v_vs_eta}
\end{figure}

The origin of the difference between these modes lies in the fact that the permeable boundary conditions act differently at degree $l=1$ and degrees $l>1$. In the case of translation, the radial velocity across the ICB is simply proportional to the topography through Eq.\;\textbf{[\ref{thermo}]}. For convection ($l>1$), the topography is also proportional to the vertical stress Eq.\;\textbf{[\ref{perm}]} and thus to $\partial v_r/\partial r$. This last relation expresses the reverse of Eq.\;\textbf{[\ref{thermo}]}: in absence of phase change, i.e. when $u_lÅ0$, the topography is maximal when the velocity across the ICB is zero. The antagonism of both relations reduces the amplitude of the radial velocity at degree $l>1$. Further, since the stress is proportional to the viscosity, the ICB becomes impermeable at low viscosity, as observed in Fig.\;\ref{v_vs_eta}.


\subsection*{Temporal evolution}
Most of the circulations shown in Fig.\;\ref{fields} are time dependent except those run with a viscosity  $\eta \geqslant10^{20}$ Pas. In a relatively limited range of viscosities, we observe periodic regimes. For instance at $10^{19}$\;Pas the energy is exchanged between degree 2 to degree 1 in an oscillating regime of period of 1.5\;Gyr. Such a period comparable with the age of the inner core age is irrelevant because inner core growth is not modeled. At $10^{18}$ Pas translation evolves pseudo-steadily while higher degrees develop chaotically. Circulation becomes completely chaotic below $10^{18}$ Pas. The time scale of variation are, in these cases, smaller than 100\;Myr and thus appears rapid at the scale of the inner core age.

\subsection*{Effect of convection velocity in the outer core}
Convection velocity $u_l$ drives material transfer through the phase change, owing to its capacity to extract or bring latent heat. It consequently drives surface radial velocity and mainly affects the translation. For instance at low viscosity ($\eta=10^{16}$ Pas), we observed that increasing $u_l$ has almost no impact on convection, except on the surface radial velocities, which are proportional to $u_l$. But neither the shape or the number of structures are affected. On the other hand, its enhancing effect on the translation is clear. This one is absent at $u_l=10^{-5}$m/s, just appears at $u_l=10^{-4}$m/s for an internal heating rate $\Phi=2.5\times10^{-8}$ W/m$^3$, with convective structure equally distributed on both hemisphere (Fig.\;\ref{fields}c). At $u_l=10^{-3}$m/s, translation is present whatever the internal heating rate, but dominates for an internal heating rate $\Phi\geqslant2.5\times10^{-9}$ W/m$^3$, as revealed by the asymmetric repartition of convective structures, 

\begin{figure}[h]
\centerline{\includegraphics[width=9cm]{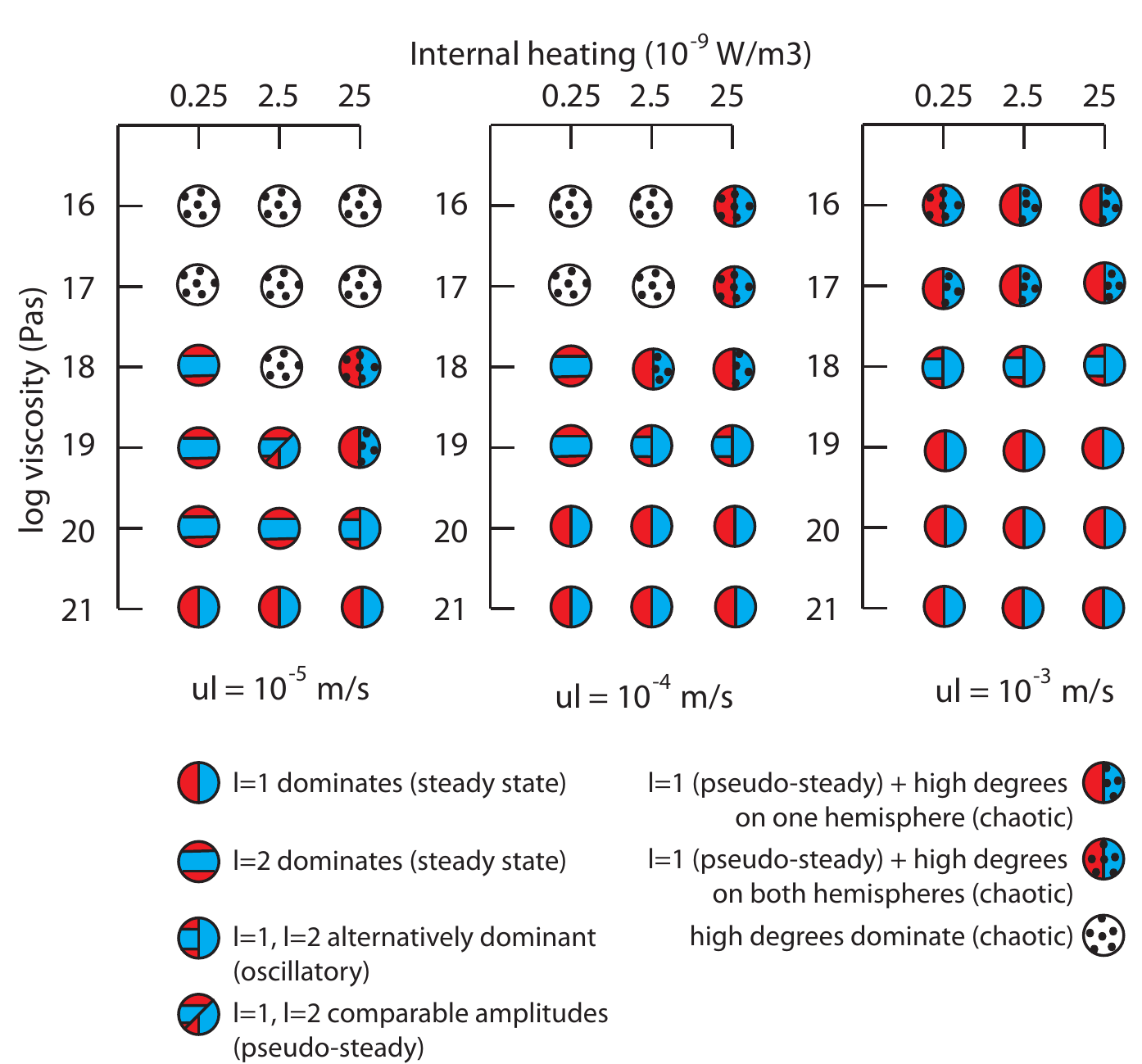}}
\caption{Summary of the convection patterns obtained in this study.}
\label{summary}
\end{figure}

Fig.\;\ref{summary} summarizes experiments performed in this study. Globally, it highlights the fact that \textit{i)} viscosity weakens the convection, \textit{ii)} the liquid velocity at the inner core surface promotes translation and \textit{iii)} internal heating promotes both convection and translation, but with a larger effect on the latter. We can see on this figure that most of the situations found at given values of $\eta$, $\Phi$ and $u_l$ are equivalent to the ones found at $\eta/10$, $\Phi/10$, and $10 u_l$

\section*{Implication for the viscosity of the inner core}

The seismic asymmetry of the inner core and particularly the correlation of velocity with attenuation of compressive wave traveling in the uppermost part of the inner core has been interpreted as the signature of iron crystal growth during their transit from a freezing side (facing Peru) to a melting side (facing Indonesia). This model requires that translation be dynamically possible. A first condition is that the inner core be in a superadiabatic state, corresponding here to $\Phi>0$ (see Eq.\;\textbf{[\ref{phi}]}). A second condition concerns the viscosity of solid iron at inner core conditions, $\eta$. If the viscosity is too low, high harmonic degree convection takes place, preventing the development of the translation. Whatever the internal heating considered here (i.e. the core cooling rate), almost pure translation prevails, above a viscosity threshold that only depends on the vigor of the outer core convection. In Fig.\;\ref{summary}, this threshold appears where the constant product of the average liquid iron velocity above the ICB $u_l$ and the viscosity exceeds $10^{16}$\;Pam. Since the most accepted value for $u_l$ lies around $10^{-4}$m/s, this threshold corresponds to $\eta=10^{20}$\;Pas.

Just below, at $\eta=10^{19}$\;Pas, harmonic degree $l=2$ establishes or alternates with the harmonic degree $l=1$ (translation) in a periodic or quasi periodic regime. Because of their time scale exceeding the age of the inner core, these situations appear irrelevant and would necessitate a modeling that takes into account inner core growth. In any case, a circulation dominated by a degree $l=2$ would not satisfy the observed seismic asymmetry. 

Translation is recovered at viscosity one order of magnitude lower, $\eta=10^{18}$\;Pas, but embedded with short wavelength convective structures that develop on the freezing side only (see cross section of Fig.\;\ref{fields}e). The resulting deformation precludes crystal growth, but may imprint a texture that could be responsible for the seismic anisotropy of the inner core. Seismic anisotropy may be subject to a hemispherical variation, with the western hemisphere displaying much stronger anisotropy than the eastern hemisphere \citep{Tanaka1997, Deuss2010}. On Fig.\;\ref{fields}e, the hotter hemisphere is not sheared by convective currents, material being simply advected toward the melting side. Annealing during this transit would erase the texture and allow crystal growth \citep{Bergman2010}. This asymmetric convection is still present at lower viscosity, $\eta\leqslant10^{17}$\;Pas, but only for values of $u_l$ ten times larger than the classically expected at the base of the outer core. Such a vigorous circulation appears in opposition with the possible presence of a stable high density layer above the ICB, produced by the fusion of inner core material depleted in light elements, for which the translation of the inner core has also been invoked \citep{Alboussiere2010}. For the expected $u_l$ value, a degree $l=1$ remains for $\eta \leqslant10^{17}$\;Pas, but short wavelength convective structures stir both hemispheres preventing any asymmetry of crystal size and orientation and the formation of a dense layer at the surface of the inner core.

In this study, the external environment was considered as homogeneous. Global circulation in the liquid core may induce variations of the thermodynamic conditions at the surface of the inner core, which are able to significantly disturb its growth rate \citep{Aubert2008, Gubbins2011}. External forcing \citep{Yoshida1996, Aubert2008, Deguen2011b} may act in conjunction with thermal convection and may stabilize degree 2 and degree 1 together.

If translation generates the seismic asymmetry of the inner core and if its anisotropy results from convection, our numerical simulations show that viscosity is comprised between $10^{18}$ and $10^{20}$ Pas for reasonable values of conductivity and age of the inner core. For viscosities lower than $10^{18}$ Pas the "sluggish" inner core is impermeable, meaning that exchanges with the outer core are small, it is deformed and has no hemispherical pattern. For viscosities higher than $10^{20}$ Pas the rigid inner core is permeable, non-deformed but hemispherical. For intermediate viscosities ($10^{18}$ Pas) particular conditions implying a boundary layer close to its Rayleigh number, allows convection to develop over a bulk translation. This configuration possibly makes thermal convection a process able to explain seismic dichotomy and large-scale variation of anisotropy.

\appendix
\section{Force balance of the inner core}

\begin{figure}
\begin{center}
\includegraphics[height=6cm]{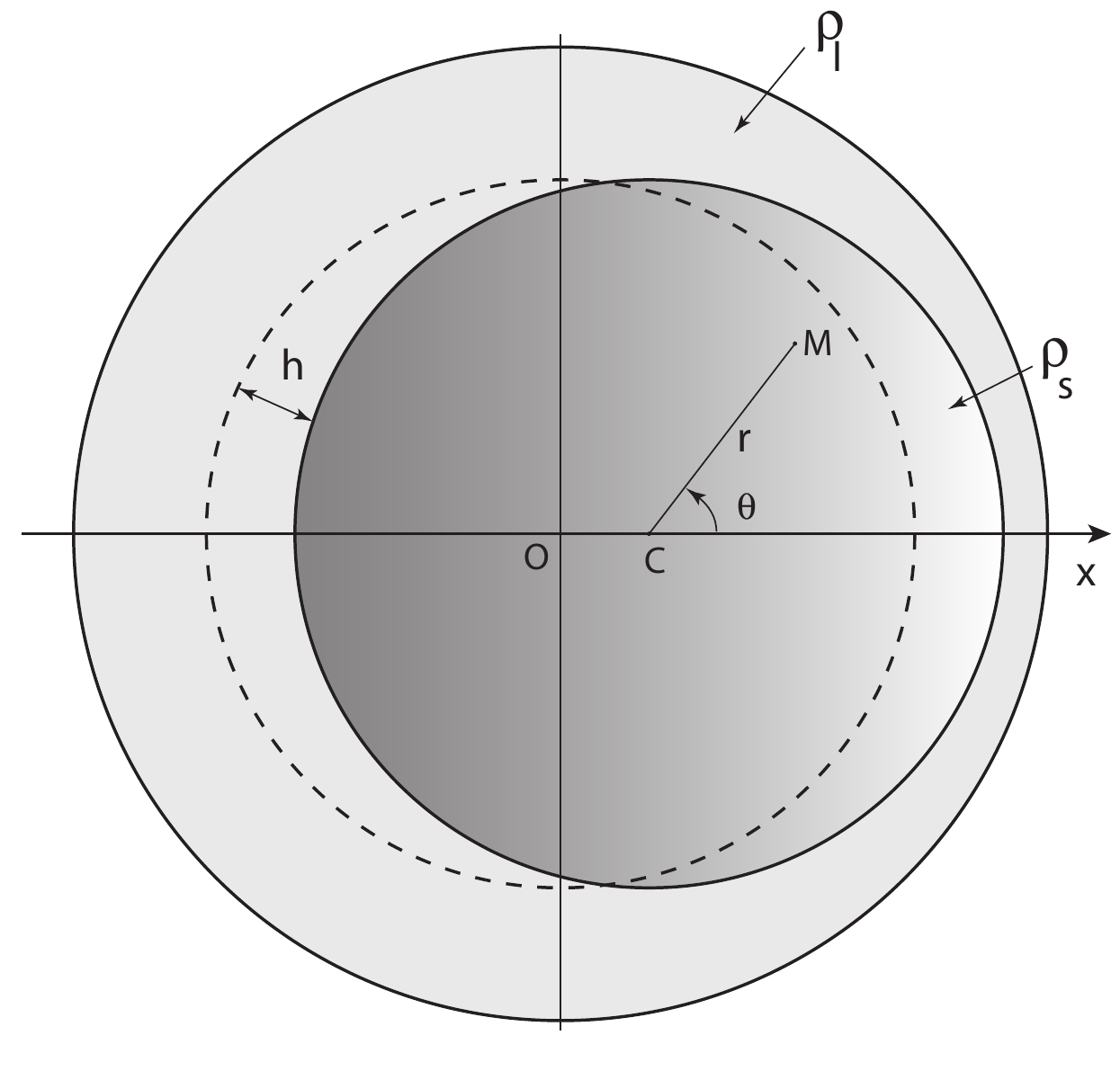}
\caption{ }
\label{Schema}
\end{center}
\end{figure}

Convection within the inner core affects the density field. The harmonic degree $l=1$ of these variations will shift the center of mass of the inner core. To visualize this shift, let us consider the inner core as a solid sphere of average density $\rho_s$, entirely covered with a liquid ocean (the outer core) of density $\rho_l$ (see Fig. \ref{Schema}). The density of the solid, marked by a variation at a spherical harmonic degree $l=1$ due to a temperature variation, is:  
\begin{equation}
\label{rho1}
\rho=\rho_s \,\left[ 1-\alpha \, \sum_{m=-1,1}\Theta_1^m(r) Y_1^m(\theta,\phi)\right],
\end{equation}
where $\Theta_1^m(r)$ are the components of degree $l=1$ of the temperature field. More simply,  we may write:
\begin{equation}
\rho=\rho_s [1-\alpha \Theta_1(r)\cos\theta],
\end{equation}
where $\theta$ is the angle made by the radius originating at the center of the figure of the solid $C$ and the direction of the axis defined by $C$ and the center of mass of the system $O$, $\Theta_1(r)$ being the radial profile of the temperature heterogeneity at degree $l=1$. The ocean surface is an equipotential of the gravity field and thus is centered on the center of mass $O$. The position of the center of mass is the one that reduces to zero the first torque of the mass anomaly that represents the inner core within the outer core, the contribution of the outer core being zero: 
\begin{equation}
\label{centre_masse}
\begin{split}
&\int_{V} \vect{OM} \rho(M)  dv =  \int_{V} \vect{OM} \rho_l  dv \; + \\
&  \int_{\text{inner core}}\vect{OM}\left[ \rho_s -\alpha \rho_s \Theta_1 (M)-\rho_l\right] dv = \vect{0}
\end{split}
\end{equation}
Introducing the radial vector $\vect{OM}=\vect{OC}+\vect{r}$ leads to: 
\begin{equation}
\label{cmg}
\begin{split}
\int\limits_{0}^{2\pi}\int\limits_{0}^{\pi}\int\limits_{0}^{R_{ic}}  (\overline{OC}+r \cos\theta)\left[ \rho_s-\rho_l-\rho_s\alpha \Theta_1(r)\cos\theta\right]  \\
r^2 \sin\theta drd\theta d\phi =0,
\end{split}
\end{equation}
and finally:
\begin{equation}
\label{OC}
\overline{OC}=\dfrac{\rho_s \alpha}{R_{ic}^3(\rho_s -\rho_l)} \int_0^{R_{ic}} r^3 \Theta_1(r) dr .
\end{equation}
The distance $\overline{OC}$ also corresponds to the amplitude $h_1$ of the harmonic degree $l=1$ of the topography of the solid sphere referenced at $O$ ; this allows to relate $h_1^m$ to $\Theta_1^m$ through the coordinate of $O$ relatively to $C$.

\section{Phase change kinetics 1D model }

A delicate point in solving this problem concerns the computation of the inner core topography. At the harmonic degree $l=1$, the topography is given by the offset position of center of mass, whereas  it results from the convection velocity, at degrees $l>1$.  The solution would consist in writing an integral boundary condition via a topography differential equation: 
\beq
\dfrac{\partial h}{\partial t} + \vect{v}_H(R_{ic}) \cdot \nabla h=v_r(R_{ic})+v_f,
\eeq
where $\vect{v}_H(R_{ic}$ is the surface horizontal velocity, $v_r(R_{ic})$ the radial velocity across the ICB and $v_f$ the solidification/melting velocity. If the kinetics of the phase change is fast, the above equation comes down to a simple balance between $v_r(R_{ic})$ and $v_f$, so that the topography can be calculated in the spectral domain.  In order to verify that adjustment of melting front is almost instantaneous compared to inner core convection time scale, we developed a 1D model of the time evolution of the melting/solidification front. 

\subsection{Model}

We solve the energy equation in which the latent heat, $L$, has been introduced as in \cite{Christensen1985} :

\begin{equation}
\label{Q}
 \small \rho C_P \frac{D T}{D t} - \rho L \frac{D \Gamma}{D t} + \alpha T \rho g v_z = k \frac{\partial^2 T}{\partial z^2} + S(z).
\end{equation}
$\Gamma$ is a function describing the solid fraction:
\begin{equation}
\begin{array}{ccc}
 \small \Gamma(\pi)= \dfrac{1}{2} (1+ \tanh (\pi)), & \textnormal{with} &  \small \pi = \dfrac{p- p_0 - \gamma T}{\delta p}.
 \end{array}
\end{equation}
$\pi$ is the pressure offset due the phase change, normalized by its pressure  thickness  $\delta p $. $\gamma$ is the Clapeyron slope of the pause change, p the pressure, $p_0$ a reference pressure. S(z), the source/sink term, corresponds to the capacity of outer core to provide or extract the phase change latent heat:
\beqn
  S(z) = - \rho C_p  (T-T_{ad}) u_l \dfrac{d\Gamma}{dz},
 \eeqn
with
   \beqn 
   \dfrac{d \Gamma}{dz}=\left( \dfrac{\partial \pi}{\partial p} \dfrac{dp}{dz}+\dfrac{\partial \pi}{\partial T}\dfrac{dT}{dz} \right) \dfrac{d \Gamma}{d \pi}.
 \eeqn
 S becomes:
\beqn
 S(z) =  \rho C_p  (T-T_{ad}) u_l \dfrac{d\Gamma}{d\pi} \dfrac{\rho g}{\delta p} \left(1+ \dfrac{\gamma}{\rho g} \dfrac{\partial T}{\partial z} \right).
\eeqn
Expanding $D \Gamma / D t $:
\beqn
\frac{D \Gamma}{D t} = \frac{d \Gamma}{d \pi} \left( \frac{\partial \pi}{\partial T}   \frac{D T}{D t} + \frac{\partial \pi}{\partial p} \frac{D p}{D t} \right),
\eeqn
and introducing the temperature dependence of latent heat:
\beqn
L= \dfrac{\gamma \delta \rho T_s}{\rho^2},
\eeqn
allows to write the specific heat $C_p'$ and the thermal expansion coefficient $\alpha'$ modified by latent heat \citep{Christensen1985}:
 \beq
 \begin{array}{ccc}
  C_p'(z) = C_p + \dfrac{T(z) \gamma^2 \delta \rho}{\rho^2 \delta p} \dfrac{d \Gamma}{d \pi} & et &   \alpha '(z) = \alpha + \dfrac{\gamma \rho}{\rho \delta p}  \dfrac{d \Gamma}{d \pi}.
  \end{array}
 \eeq
Finally conservation of heat (\ref{Q}) becomes:
 \beqn
 \label{eq:Q2}
 \rho C_p' \left[\dfrac{\partial T}{\partial t} + v_z \dfrac{\partial T}{\partial z} \right]+ \alpha ' \rho g v_z T = k \dfrac{\partial^2 T}{\partial z^2} + S(z).
 \eeqn
 
We use a Backward Euler scheme to solve differential equation (\ref{eq:Q2}). Boundary conditions are differently defined when material enters or leave the box. Solid Iron (the inner core) is supposed to occupy the bottom half of the domain. In the case of solidification ($v_z<0$), the top temperature is ascribed to the outer core temperature, which is $T_{top}=T_{ad}$. At the bottom, we impose the temperature gradient to be equal to the adiabatic gradient. In the case of melting ($v_z > 0$),  the bottom temperature is set to the adiabatic temperature plus an anomaly resulting from internal heating: $T_{top} = T_{ad} + \delta T$. At the top, we impose the temperature gradient to be equal the adiabatic gradient.

\subsection*{Results}

We performed five simulations varying the vertical velocities $v_z$ --- supposed to be the radial velocity across the Inner core Bounadry (ICB)---, the convective velocity within the liquid outer core just above the ICB and the temperature within the inner core.  Table \ref{tab} summarizes the characteristics of these experiments. 
\begin{table}[b]
\caption{Numerical experiment characteristics. $v_z$ is the vertical velocity, $u_l$ convection velocity inthe outer core, h the topography at thermal equilibrium.}
\centering
\begin{tabular}{ccccc}
\\
\hline
Exp. & $v_z$ (cm/yr) & $u_l$ (m/s)  & Initial T(z) &  (h) (m) \\ \hline
1                   &          2.5             & $1 \times 10^{-4}$ & Adiabat & 151.46 \\   
2                   &         -2.5             & $1 \times 10^{-4}$ & Adiabat & -151.46  \\ 
3                   &          2.5             &	$1 \times 10^{-4}$ & Adiabat+1K & 151.46 \\  
4                   &          2.5             & $ 3 \times 10^{-4}$ & Adiabat & 50.49 \\ 
5                   &          1.25          & $1 \times 10^{-4}$  & Adiabat & 75.73 \\ \hline
\end{tabular}
\label{tab}
\end{table}

\begin{figure}
\centering
\includegraphics[width=8 cm]{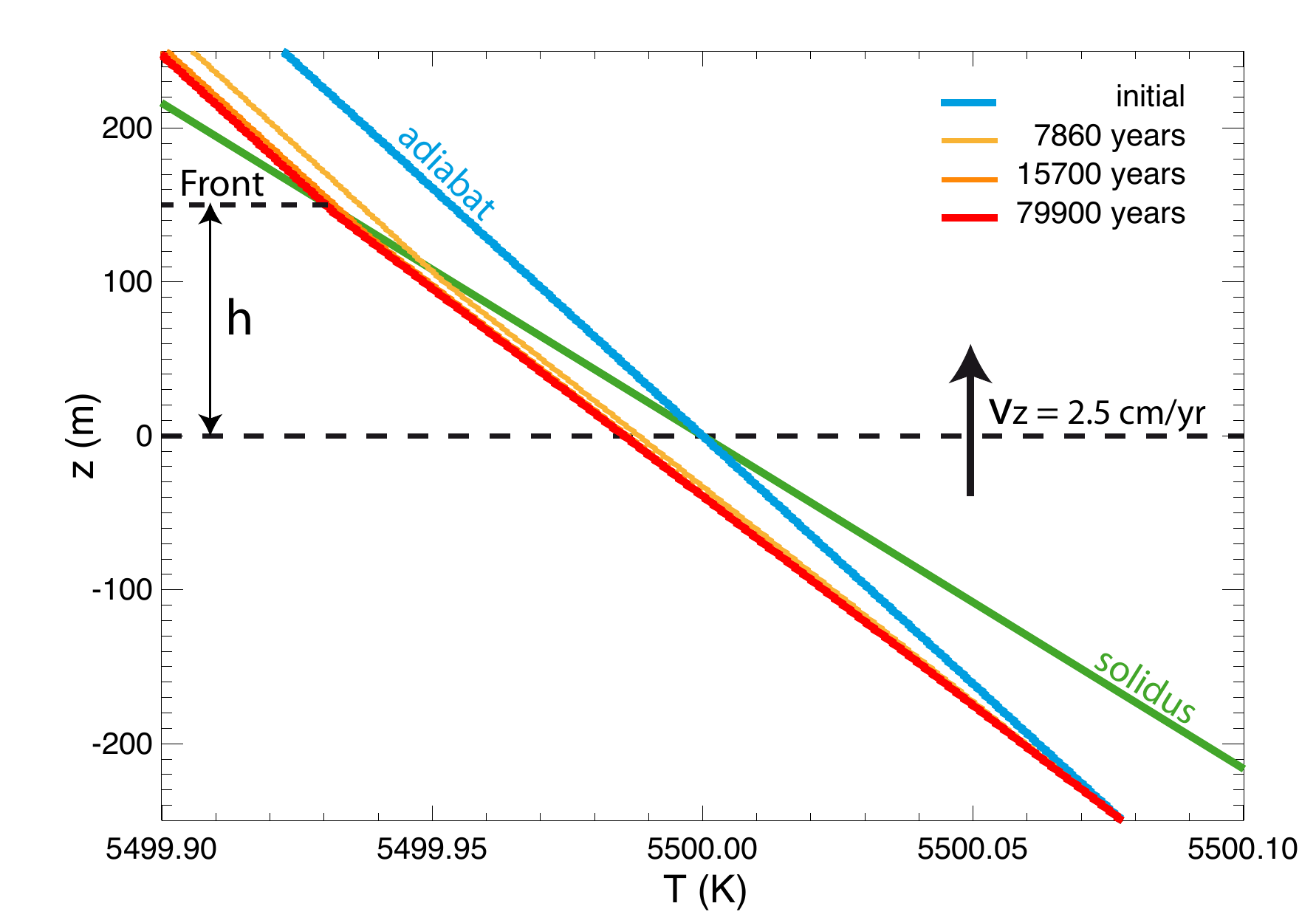}
\caption{\textit{Temperature profile evolution for simulation 1.Initial temperature profile is in blue thick line. Iron melting curve is in green thick line. Temperature profile at thermal equilibrium is red. Shaded orange tons correspond to intermediate profiles.}}
\label{exp1}
\end{figure}

Fig.\;\ref{exp1} show the time evolution of  temperature profiles in case of melting by decompression ($v_z>0$). We see that the consumption of latent heat cools the inner core by a few hundredths of kelvin, shifting up the phase change by a hundred meters. This state is reached in less that $10^4$ years, which is shorter than time scale involved in convective process. A symmetric situation is observed in the case of solidification when $v_z <0$. The time necessary to reach a steady state depends on the difference between the temperature of the inner core material and the adiabat. This time is four time larger for 1K of difference. On the other hand, we find, as expected, that the topography reached is non dependent on this parameter and that it is just proportional to $v_z$ and to $1/u_l$

Equilibrium specific time, on the order of $10^4$ to $10^5$ years, remains short enough when compared to convection time scales to consider thermal equilibrium of the topography as instantaneous.

\begin{figure}
\centering
 \includegraphics[scale=0.38]{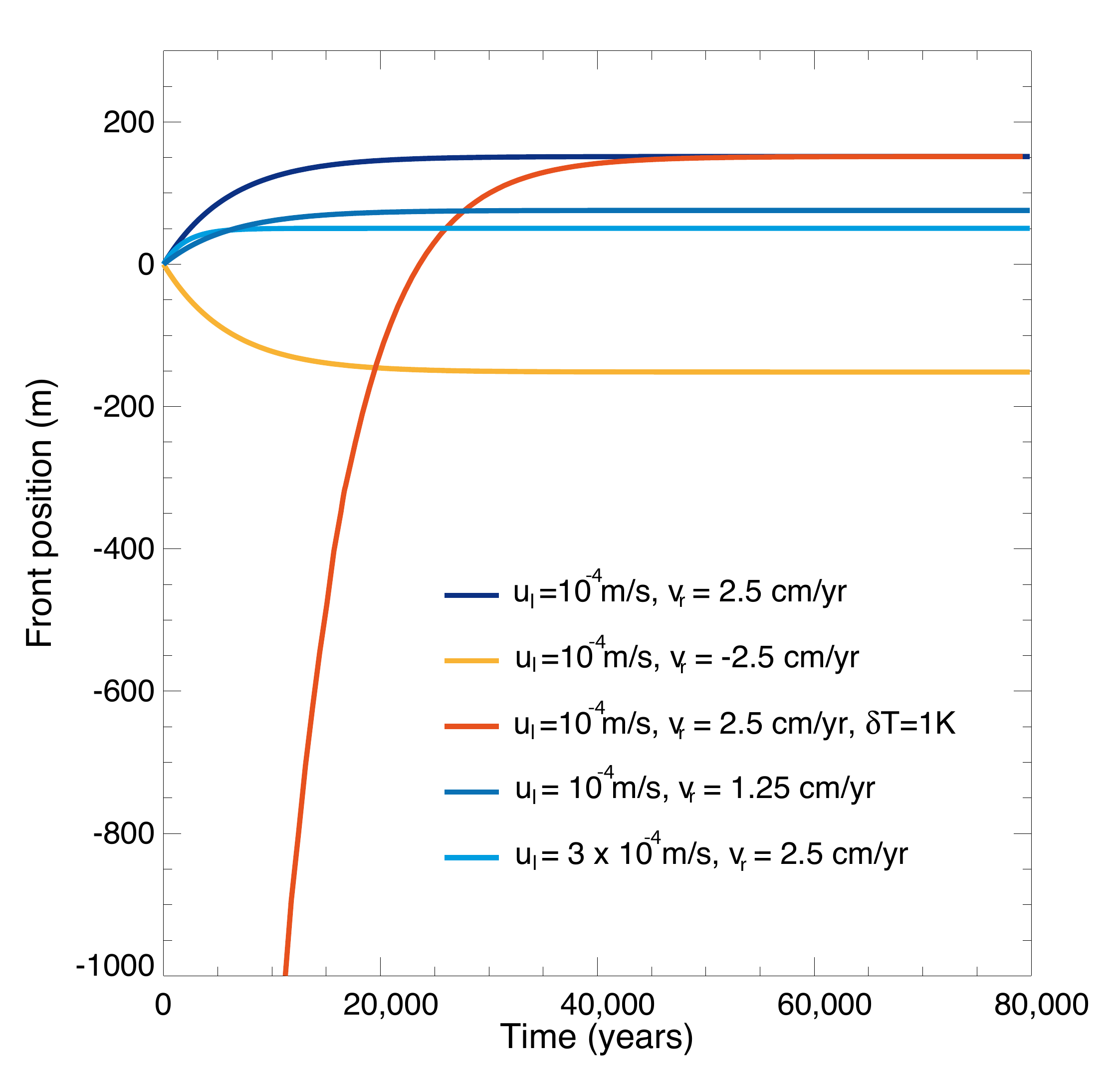}
 \caption{\textit{Topography evolution for the five simulations.}}
  \label{h_t}
\end{figure}

\bibliographystyle{elsarticle-harv}
\bibliography{biblio}

\end{document}